# A FRAMEWORK OF TEXT-DEPENDENT SPEAKER VERIFICATION FOR CHINESE NUMERICAL STRING CORPUS

Litong Zheng, Feng Hong, Member, IEEE, Weijie Xu, Wan Zheng

*Abstract*—The Chinese numerical string corpus, serves as a valuable resource for speaker verification, particularly in financial transactions. Researches indicate that in short speech scenarios, text-dependent speaker verification (TD-SV) consistently outperforms text-independent speaker verification (TI-SV). However, TD-SV potentially includes the validation of text information, that can be negatively impacted by reading rhythms and pauses. To address this problem, we propose an end-to-end speaker verification system that enhances TD-SV by decoupling speaker and text information. Our system consists of a text embedding extractor, a speaker embedding extractor and a fusion module. In the text embedding extractor, we employ an enhanced Transformer and introduce a triple loss including text classification loss, connectionist temporal classification (CTC) loss and decoder loss; while in the speaker embedding extractor, we create a multi-scale pooling method by combining sliding window attentive statistics pooling (SWASP) with attentive statistics pooling (ASP). To mitigate the scarcity of data, we have recorded a publicly available Chinese numerical corpus named SHALCAS22A (hereinafter called SHAL), which can be accessed on OpenSLR. Moreover, we employ data augmentation techniques using Tacotron2 and HiFi-GAN. Our method achieves an equal error rate (EER) performance improvement of 49.2% on Hi-Mia and 75.0% on SHAL, respectively.

*Index Terms*—Speaker verification, speaker embeddings, transformer, attentive statistic pooling, deep neural networks.

## I. INTRODUCTION

TEXT-DEPENDENT speaker verification (TD-SV) requires matching the text content in both enrollment and verification speech, resulting in fixed phonetic information. This leads to better performance metrics like equal error rate (EER) compared to text-independent speaker verification (TI-SV) [1]. TI-SV, being text-agnostic, offers flexibility for various scenarios. However, TD-SV faces challenges like collecting sufficient data for different text requirements, limiting its practical use. In speaker verification (SV), networks like X-vector, ECAPA-TDNN, Conformer, and others [2-6] have shown excellent results in both TD-SV [7-9] and TI-SV [10-12]. Yet, TD-SV has two key challenges: (1) Compared with TI-SV, the dataset [13] is limited due to fixed-text requirements. (2) Domain mismatch [1,13], especially different text which negatively effects the performance of speaker verification.

Much useful work has been done on TD-SV. Qin *et al.* augmented the TD-SV dataset through text to speech (TTS) with Tacotron2 and voice conversion (VC) [17][18] to mitigate the lack of data. Qian *et al.* [12] enhanced system performance by optimizing text-related gradients before pooling operations to resolve domain mismatch. They also introduced a speaker-text factorization network [14], which first encode the speech as a text-independent speaker and text-dependent text embedding and then recombined them into a single embedding. Han *et al.* [15] improved ECAPA-TDNN for short-segment SV with a time-domain multi-resolution encoder. Mak *et al.* [16] used weight space ensemble for SV challenges, particularly domain mismatch, outperforming traditional methods.

We aim to apply TD-SV to financial payments identity verification scenarios, which requires better EER performance compared with other applications. In the verifying phase, the validation word, which is a long text of random numbers, appears randomly such as "8 1 7 3 2 5 9 6 0 4". When calculating the similarity score, we segment and sequence the random digital string verification utterances to make the order of the verified voice the same as the designed ones. In such case, it is seen that when the length of the text is long, the model exhibits limited sensitivity to the text's order, resulting in a significant performance degradation. Thus, an end-to-end TD-SV framework has been introduced. Our main contributions are as follows:

(1) We established a comprehensive Chinese numerical string dataset and validated the efficacy of TTS augmentation in enriching its variability, including the incorporation of nuanced features like stopping rhythms.

(2) Our dual-ended network efficiently extracts both text and

This work was in part supported by the National Natural Science Foundation of China under Grant 61971371. (Corresponding author: Feng Hong.)
Litong Zheng is with Shanghai Acoustics Laboratory, Chinese Academy of Sciences, Shanghai, China, and University of Chinese Academy of Sciences, Beijing, China (e-mail: zhenglitong21@mails.ucas.ac.cn).
Feng Hong is with Shanghai Acoustics Laboratory, Chinese Academy of Sciences, Shanghai, China (e-mail: hongfeng@mail.ioa.ac.cn).

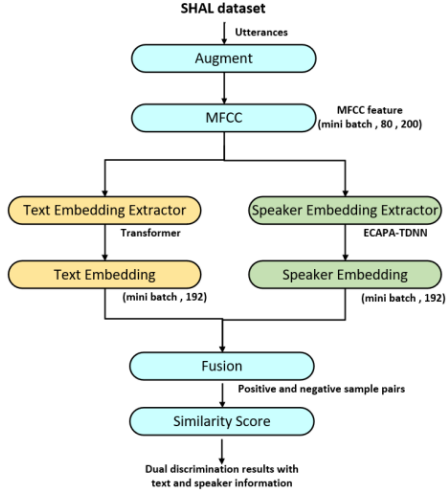

Fig. 1. The overall framework diagram of end-to-end TD-SV system. Above are data augmentation and MFCC feature extraction. On the left and right sides, we have Transformer and ECAPA-TDNN extracting text and speaker embedding, Finally, these embeddings are fused.

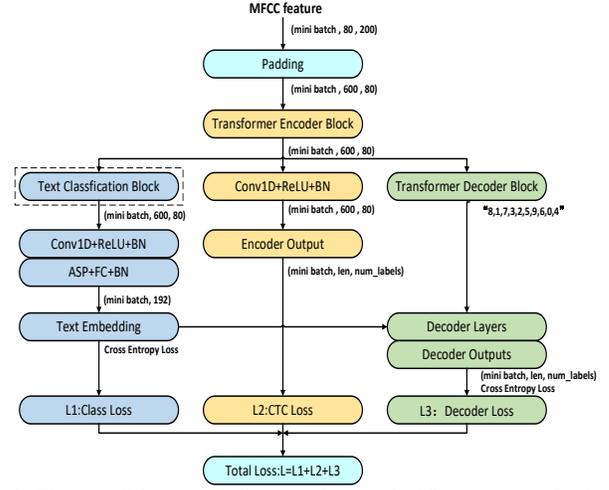

Fig. 2. The overall framework diagram of text embedding extractor. On the left, we have classification loss; in the middle, there is the CTC Loss; on the right, we have decoder loss.

speaker embeddings from speech. Notably, the text embedding extractor incorporates an enhanced Transformer architecture, enhancing the network's sensitivity to textual sequencing.

(3) A novel pooling method named Sliding Window Attentive Statistics Pooling (SWASP), is introduced and integrated with the original Attentive Statistics Pooling (ASP) method that is initially employed in ECAPA-TDNN, to comprehensively address phonetic variability. Recognizing ASP's limitation in adequately capturing temporal dynamics, particularly in fixed-text corpora, we leverage SWASP to achieve temporal encoding and compression via sliding window pooling within the temporal dimension.

(4) We integrate text and speaker embeddings via diverse fusion methods such as addition, multiplication, or CNN fusion [21], drawing inspiration from the fusion methods in the Spoofing Aware Speaker Verification Challenge (SASVC2022) [20].

## II. 2. END-TO-END TD-SV FRAMEWORK

As shown in Fig.1, our end-to-end TD-SV framework involves several parts: (1) In the top blue sections, data augmentation including TTS and speed disturbance and MFCC feature extraction are performed. (2) In the left-hand yellow section, the text embedding extractor based on Transformer is employed to deal with text information. (3) In the right-hand green section, the speaker embedding extractor based on ECAPA-TDNN is employed. (4) In the bottom blue section, the fusion module combines the embeddings obtained from above.

### A. Text embedding extraction based on Transformer

Motivated by the work of Qian [14] and speech recognition engine We-Net [22], the text embedding extractor based on a single Transformer is proposed to fully capture the phonetic information in text, as shown in Fig.2. The extractor includes positional encoding module, encoder and decoder following the structure of standard Transformer. The text representation yields three distinct outputs, each associated with a unique loss function as follows:

(1) The left part is classification loss. We introduce a text classifier of a simple CNN block after the encoder. Following classification, ASP is applied resulting in a 192-dimensional output serving as text embedding. This is followed by further dimension reduction through a linear layer, with the output dimension matching the number of text labels, assuming a finite set of text labels for classification. Finally, we obtained classification loss $\mathcal{L}_1$:

$$E_{text} = Pooling(BN(\mathrm{Re}LU(Conv1D(X)))), \quad (1)$$

$$\mathcal{L}_1 = CE(BN(FC(E_{text}))), \quad (2)$$

where $X$ is the output of encoder, and $E_{text}$ is the text embedding. The total CNN block consists of Conv1D, ReLU, BN and ASP pooling.

(2) The middle part in yellow sections is CTC loss. The output of encoder passes a three-layer CNN before calculating CTC loss. The CTC loss function addresses alignment issues between labels and predictions in neural networks, enabling the model to handle speech with varying pause rhythms. Consequently, CTC loss $\mathcal{L}_2$ can be represented by:

$$\mathcal{L}_2 = CTC(BN(\mathrm{Re}LU(Conv1D(X))*3)), \quad (3)$$

where "*3" represents three CNN blocks and CE stands for cross-entropy loss function.

(3) The right part in green sections is decoder loss. Since the phoneme information incorporates rhythms and pauses, we use the output of text classifier as phoneme label during decoding. This facilitates faster convergence and makes decoded results closely resemble the true text sequence. This yields the decoder loss $\mathcal{L}_3$:

$$\mathcal{L}_3 = CE(Y), \quad (4)$$

where $Y$ is the output of decoder.

The overall loss size is represented as $\mathcal{L}_{total}$:

$$\mathcal{L}_{total} = \alpha\mathcal{L}_1 + \beta\mathcal{L}_2 + \gamma\mathcal{L}_3, \quad (5)$$

where α, β and γ represent the weighting coefficients of the three losses, respectively. In our work, we emphasize text classification capability, so we set the three weights of $\mathcal{L}_{total}$ to 0.6, 0.2 and 0.2, respectively. In general, the architecture of our text embedding extractor is similar to a speech recognition engine, with the difference that we introduce an additional text classifier, which makes the text embedding accurate to some extent.

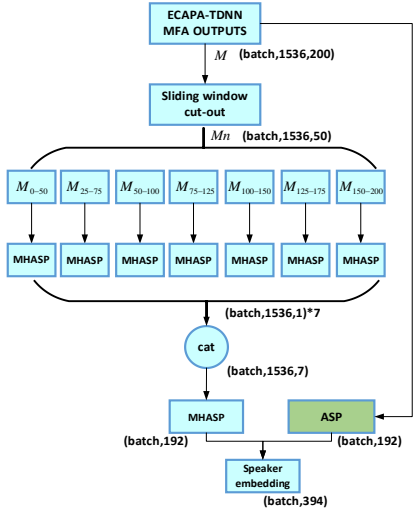

Fig. 3. The diagram of Sliding Window Attentive Statistics Pooling (SWASP). Firstly, a sliding window is used to divide the output of MFA into several segments. Afterwards, we apply MHSSP to each segment, and then perform MHASP again on the concatenated outputs. This gives the result for SWASP. Simultaneously, we also perform a traditional ASP, and the speaker embedding is composed of these two parts together.

*B. Speaker embedding extraction and SWASP*

The overall computational diagram of SWASP is illustrated in Fig.3. The ECAPA-TDNN multi-layer feature aggregation(MFA) output, denoted as $\mathcal{M}$, has a size of $[batch, 1536, 200]$. By utilizing a window with a length of 50 and a stride of 25, $\mathcal{M}$ can be divided along the time dimension into different parts like $[\mathcal{M}_{0-50}, \mathcal{M}_{25-75}, \mathcal{M}_{50-100}, \mathcal{M}_{75-125}, \mathcal{M}_{125-150}, \mathcal{M}_{150-175}, \mathcal{M}_{175-200}]$, each with a size of $[batch, 1536, 50]$. After applying the multi-head attentive statistics pooling (MHASP) to each part, we obtain an output of size $[batch, 1536]$. All the outputs are concatenated and MHASP is employed once more to yield the final SWASP result, which has a size of $[batch, 192]$. On the right side of Fig.3, the traditional ASP is also applied and together with the results of SWASP compose the speaker embedding.

In each MHASP, three linear transformations are used to obtain the matrices $Q$, $K$ and $V$ [19]. After the allocation of multi-head attention, we get vector $A$ as follows:

$$Q, K, V = Linear1(X), Linear2(X), Linear3(X) \quad (6)$$

$$A(Q, K, V) = Softmax(QK^T / \sqrt{d_k})V \quad (7)$$

$$\alpha = Softmax(\omega_2(tanh(\omega_1 A + b_1) + b_2, dim = 2) \quad (8)$$

$$\mu = \sum_t^T \alpha_t X_t, \sigma^2 = \sum_t^T \alpha_t X_t \odot X_t - \mu \odot \mu \quad (9,10)$$

where $X$ is the input of MHASP, and $A$ is the result obtained after multi-head self-attention, $\alpha$ represents the weights obtained by ASP for $A$. $\mu$ and $\sigma^2$ are the first-order and second-order statistics obtained from ASP, respectively. $\odot$ is Hadamard Product.

*C. Dataset and data augmentation*

SHAL dataset comprises approximately 72.3 hours of audio with 46,583 files in 44.1kHz, 16-bit PCM-WAV format. This dataset is specifically tailored to speakers aged 10-40, with gender balance being a key consideration. Table 1 summarizes the dataset. We selected 60 individuals, each contributing 25 samples for each text type.

TABLE I
SHALCAS22A Recording Information Statistics

| Text label | Text content | Speech duration |
|---|---|---|
| d001 | 8-1-7-3-2-5-9-6-0-4 | 6s |
| d002 | 8-1-7-3\|2-5-9-6\|-0-4 | 4s |
| d003 | 8-1-7\|-3-2-5\|-9-6-0\|-4 | 4s |
| d004 | 8-1\|-7-3\|-2-5\|-9-6\|-0-4 | 4s |
| d005 | 9-4-0-5\|3-7-2-6\|-8-1 | 4s |
| d006 | 9-4-0\|-5-3-7\|-2-6-8\|-1 | 3s |

For data augmentation, we use Tacotron2 and HiFi-GAN [17] to synthesize more Chinese numerical corpus. Initially, we perform pretraining on 10,000 female voices from AISHELL to get pretrained model. Subsequently, we apply transfer learning on the speech of 60 speakers in SHAL, and a total of 60 personalized TTS models are obtained. Finally, we conduct TTS with different pause rhythms on our trained model. Meanwhile, all speech audio (including the original SHAL and TTS enhancement) is performed 0.9x and 1.1x speed perturbations. Through these augmentation techniques, the total dataset size increases sixfold.

## III. EXPERIMENTAL SETUPS

Our experiments are as follows: (1) Experiments on different data augmentation method. The network is ECAPA-TDNN, which is trained on VoxCeleb2 with 5994 speakers. The evaluation datasets include VoxCeleb1-O, the original SHAL dataset, and various combinations of SHAL and augmented data with speed perturbation and TTS enhancement. (2) Experiments on different pooling methods in ECAPA-TDNN. The data use for training and evaluation contains VoxCeleb1-O, Hi-Mia (near-field noise-free dataset) and original SHAL (d002 and d005). (3) Experiments on different SWASP hyperparameters settings. The network is still ECAPA-TDNN, while training and evaluating is performed on Hi-Mia. (4) Experiments on different fusion strategy of text and speaker embeddings. All experiments are conducted without noise or reverberation. Except for (1), we divide all datasets into training and evaluation sets according to 8:2. Each experiment starts with model pretraining on Voxceleb2, followed by fine-tuning and testing for their respective task.

Drawing inspiration from advanced techniques in Text-Independent Speaker Verification (TI-SV), such as the inter top-K penalty for optimizing hard samples proposed by Zhao *et al.* [25] and large margin fine-tuning proposed by T. J [26], we adopt the Adam optimizer with an initial learning rate of 0.001, decayed by 3% every epoch. The loss function used is AAM-Softmax with margin and scale set to 30 and 0.2 [25], respectively. In the text embedding extractor, both the encoder and the decoder consist of 4 blocks, each with 4 heads and dimensions of 64 of Q, K, V. The speaker embedding extractor utilizes an attention mechanism with 2 heads and only 1 block.

## IV. EXPERIMENTAL RESULTS AND ANALYSIS

Table 2. shows ECAPA-TDNN model performance trained on VoxCeleb2 and evaluated with various datasets. We focus on the performance on SHAL and its augmented datasets. Experiment indicates that EER can be significantly reduced by adding augmented data, which improves speech quality.

TABLE II
Experimental result of different evaluation dataset. 'Ori-SHAL' represents the original SHAL dataset. 'Speed' denotes the dataset after data perturbation and 'TTS' stands for the augmented dataset synthesized by Tacotron2.

| Evaluation dataset | Sample Nums | EER(%) | minDCF |
|---|---|---|---|
| VoxCeleb1-O | 37611 | 1.09 | 0.078 |
| Ori-SHAL | 7140 | 0.69 | 0.050 |
| Ori-SHAL + Speed | 28800 | 1.44 | 0.118 |
| Ori-SHAL + TTS | 19200 | 1.93 | 0.245 |
| Speed + TTS | 28800 | **0.27** | **0.025** |
| Ori-SHAL + Speed+ TTS | 57600 | 1.72 | 0.171 |

TABLE III
Experimental result of different pooling strategy. 'A' represents ASP, while 'M' represents single MHASP and 'S' represents single SWASP.

| System | VoxCeleb1-O | | Hi-Mia | | SHAL | |
|---|---|---|---|---|---|---|
| | EER(%) | minDCF | EER(%) | minDCF | EER(%) | minDCF |
| A-base | **1.09** | **0.078** | 0.63 | 0.046 | .0.48 | 0.025 |
| M | 1.36 | 0.097 | 1.11 | 0.054 | 0.56 | 0.028 |
| S | 1.35 | 0.095 | 0.79 | 0.070 | 1.67 | 0.100 |
| A+S | 1.18 | 0.083 | **0.32** | **0.033** | **0.12** | **0.011** |
| M+S | 1.29 | 0.088 | 0.84 | 0.071 | 0.35 | 0.056 |
| A+M+S | 1.35 | 0.096 | 0.68 | 0.046 | 0.55 | 0.036 |

Table 3. shows the performance of different pooling methods in ECAPA-TDNN. Each model is fine-tuned with a new dataset after being pretrained on VoxCeleb2 and evaluated independently. As anticipated, experiments reveal that SWASP, designed for text-dependent tasks, underperformscompared to the baseline (ASP) when dealing with text-independent tasks (VoxCeleb1-O). Utilizing a single MHASP or SWASP results in a weakened overall representation capability, leading to a slight performance decrease. The best performance is attained through the combined usage of ASP and SWASP, resulting in a remarkable 49.2% and 75.0% improvement in performance on Hi-Mia and SHAL, respectively. This combination offers text-independent global representation capability while maintaining some text-dependent local representation capability, therefore the best performance is achieved. Furthermore, it can be concluded from Table 3. that using a single MHASP or SWASP may lead to performance degradation due to potential issues such as overfitting. However, an increase in performance is observed when combining them with ASP to form a multi-scale pooling.

Table 4 illustrates the results of SWASP hyperparameters experiments by varying window length and stride. The model is pretrained on Voxceleb2, followed by transfer learning and fine-tuning on Hi-Mia, and finally evaluate on Hi-Mia. The results indicate that the optimal performance is achieved with a window length (w) of 50 and a stride (s) of 25. In practical experiments, we observe that an excessively short window length can reduce the computation speed, while an overly long one can increase the pressure on memory resources.

TABLE IV
Experimental result of different SWASP hyperparameter settings. 'w' denotes window length and 's' represents stride.

| System | EER(%) | minDCF |
|---|---|---|
| w=25, s=25 | 0.48 | 0.038 |
| w=50, s=50 | 0.36 | 0.041 |
| w=75, s=75 | 0.39 | 0.042 |
| w=100, s=100 | 0.35 | 0.041 |
| w=50, s=10 | 0.33 | 0.041 |
| w=50, s=20 | 0.36 | 0.039 |
| w=50, s=25 | **0.32** | **0.033** |
| w=50, s=30 | **0.32** | 0.040 |
| w=50, s=40 | 0.37 | 0.042 |

TABLE V
Experimental result of different fusion strategy on different models. 'A-base', 'S' and 'A+S' are the same in Table 3. 'A' represents ECAPA-TDNN with single ASP fine-tuned on SHAL.

| System | Addition | Multiplication | CNN fusion |
|---|---|---|---|
| | EER(%) | EER(%) | EER(%) |
| A-Vox2-base | 1.11 | 1.08 | 8.33 |
| A-SHAL | 0.78 | 0.69 | 18.33 |
| S-SHAL | 1.67 | 1.67 | 8.89 |
| (A+S)-SHAL | **0.20** | **0.20** | 10.04 |

Table 5shows the performance of the fusion module, where text embedding and speaker embedding are combined, allowing the system to consider both speaker identity and text categories. Following the SASV2022 approach, two types of text are used, and three fusion strategies are proposed: embedding addition, embedding multiplication and embedding CNN fusion. We maintain consistency by using the same text embedding across different speaker embeddings, obtained from the models listed in Table 3 (A-base, S, and A+S), except for a newly introduced model with ASP fine-tuned on SHAL. As previously, the training set to evaluation set ratio remains 8:2. Experimental findings indicate that the embedding multiplication strategy achieves the best performance, yielding an EER of 0.20% on our proposed ECAPA-TDNN model with ASP and SWASP. Conversely, the CNN fusion strategy exhibits notable instability in comparison.

## V. CONCLUSION

We present an efficient framework for TD-SV utilizing the Chinese numerical string corpus, incorporating data augmentation, text and speaker embedding extraction, and embedding fusion. Our novel multi-scale pooling method, SWASP, significantly enhances performance by 49.2% on Hi-Mia and 75.0% on SHAL datasets. Additionally, by introducing an embedding fusion strategy, we attain the optimal performance with a 0.20% EER on SHAL. Moving forward, our efforts will concentrate on enhancing the versatility of our method and optimizing model size for improved efficiency.